# Implications of the hybrid epithelial/mesenchymal phenotype in metastasis

Mohit Kumar Jolly[1,2], Marcelo Boareto[1,8], Bin Huang[1,3], Dongya Jia[1,6], Mingyang Lu[1], Jose' N Onuchic[1,3,4,5,*], Herbert Levine[1,2,4,*], Eshel Ben-Jacob[1,8,*]

[1]Center for Theoretical Biological Physics; Departments of [2]Bioengineering, [3]Chemistry, [4]Physics and Astronomy, and [5]Biosciences, [6]Graduate Program in Systems, Synthetic and Physical Biology, Rice University, Houston, TX 77005-1827, USA

[6]School of Physics and Astronomy and The Sagol School of Neuroscience, Tel-Aviv University, Tel-Aviv 69978, Israel

[8]Institute of Physics, University of Sao Paulo, Sao Paulo 05508, Brazil

[*]Authors to whom correspondence should be sent: Jose' N Onuchic (Email: jonuchic@rice.edu), Herbert Levine (Email: herbert.levine@rice.edu); Eshel Ben-Jacob (Email: eshel@rice.edu)

**ABSTRACT**
Understanding cell-fate decisions during tumorigenesis and metastasis is a major challenge in modern cancer biology. One canonical cell-fate decision that cancer cells undergo is Epithelial-to-Mesenchymal Transition (EMT) and its reverse Mesenchymal-to-Epithelial Transition (MET). While transitioning between these two phenotypes – epithelial and mesenchymal, cells can also attain a hybrid epithelial/ mesenchymal (i.e. partial or intermediate EMT) phenotype. Cells in this phenotype have mixed epithelial (eg. adhesion) and mesenchymal (eg. migration) properties, thereby allowing them to move collectively as clusters of Circulating Tumor Cells (CTCs). If these clusters enter the circulation, they can be more apoptosis-resistant and more capable of initiating metastatic lesions than cancer cells moving individually with wholly mesenchymal phenotypes, having undergone a complete EMT.

Here, we review the operating principles of the core regulatory network for EMT/MET that acts as a 'three-way' switch giving rise to three distinct phenotypes – epithelial, mesenchymal and hybrid epithelial/mesenchymal. We further characterize this hybrid E/M phenotype in terms of its capabilities in terms of collective cell migration, tumor-initiation, cell-cell communication, and drug resistance. We elucidate how the highly interconnected coupling between these modules coordinates cell-fate decisions among a population of cancer cells in the dynamic tumor, hence facilitating tumor-stroma interactions, formation of CTC clusters, and consequently cancer metastasis. Finally, we discuss the multiple advantages that the hybrid epithelial/mesenchymal phenotype have as compared to a complete EMT phenotype and argue that these collectively migrating cells are the primary 'bad actors' of metastasis.

**Introduction**

Despite remarkable progress in charting the hallmarks of cancer, understanding the cell-fate decisions during tumor initiation, progression, dormancy, and relapse is a major challenge in modern oncology (1). These dynamic decisions enable the tumor cells to tolerate therapeutic assaults such as chemotherapy or radiation; adapt to common micro-environmental stress that they face during cancer progression such as hypoxia, nutrient deprivation, and inflammation; and complete their 'metastasis-invasion cascade' to seed tumors in distant organs at early stages; thereby posing unpleasant surprises in the clinical trials.

Recently, there has been rapid progress in characterizing these cell-fate decisions or cellular plasticity by mapping the underlying regulatory networks associated with the tumor-stroma ecosystem such as epithelial-mesenchymal plasticity, dedifferentiation of cancer cells to Cancer Stem Cells (CSCs), drug resistance, cell senescence, metabolic reprogramming, response to hypoxia, and tumor angiogenesis. Cell-fate determination in these examples involve changes in expression of various transcription factors (TFs), miRNA (miRs), and epigenetic regulators that govern the underlying regulatory networks and consequently generate genome-wide distinct expression patterns of genes and proteins corresponding to a particular cell fate.

An archetypical example of cell-fate decisions or cellular plasticity during tumor progression is the transition between epithelial and mesenchymal phenotypes – Epithelial to Mesenchymal Transition (EMT) and its reverse MET. EMT marks the first step of 'invasion-metastasis cascade' where epithelial cells of the primary tumor lose their cell-cell adhesion and apico-basal polarity, and gain the ability to migrate individually and invade basement membrane and blood vessels. Upon intravasation, these cells stay in the bloodstream as Circulating Tumor Cells (CTCs), until they exit at some distant organs to seed micrometastases. During seeding, they undergo the reverse of EMT – MET – to regain their epithelial characteristics and form secondary tumors or macrometastases, thereby completing their 'metastasis-invasion cascade'. Therefore, EMT and MET enable solid tumors, over 90% of which are epithelial in nature (carcinomas)(2), to disseminate and colonize distant organs. However, EMT and MET are not exclusive to cancer, rather they play crucial roles in organogenesis during embryonic development, and wound healing or tissue regeneration where they are regulated tightly, but cancer cells 'hijack' this developmental process for metastasis (3,4) – the cause of nine out of ten cancer-related deaths (5).

Importantly, EMT and MET, whether in physiological or pathological contexts, are not binary processes (3,4). Some cells can attain a hybrid epithelial/mesenchymal (E/M) phenotype, also referred to as partial or intermediate or incomplete EMT phenotype (6–8). In fact, many carcinoma cells may metastasize without completely losing an epithelial morphology and/or completely attaining mesenchymal traits (2,9). Cells in the hybrid E/M phenotype have both epithelial (cell-cell adhesion) and mesenchymal (migration) traits, hence allowing collective cell migration, as seen during migration of multicellular aggregates in ECM (2) and clusters of CTCs in bloodstream of breast, lung and prostate cancer patients (10–12). Cells in CTC clusters co-express epithelial and mesenchymal markers (13), can exit the bloodstream more efficiently (14), are apoptosis-resistant, and can be up to 50-times more metastatic than individually migrating CTCs (15). Therefore, the ability of metastatic cells to attain this hybrid E/M phenotype, rather than a complete EMT phenotype, poses a higher metastatic risk in patients (16).

Here, we focus our review on elucidating how cells attain this phenotype, characterizing this hybrid E/M phenotype, and discussing why cells in this phenotype are the primary 'bad actors' of cancer metastasis.

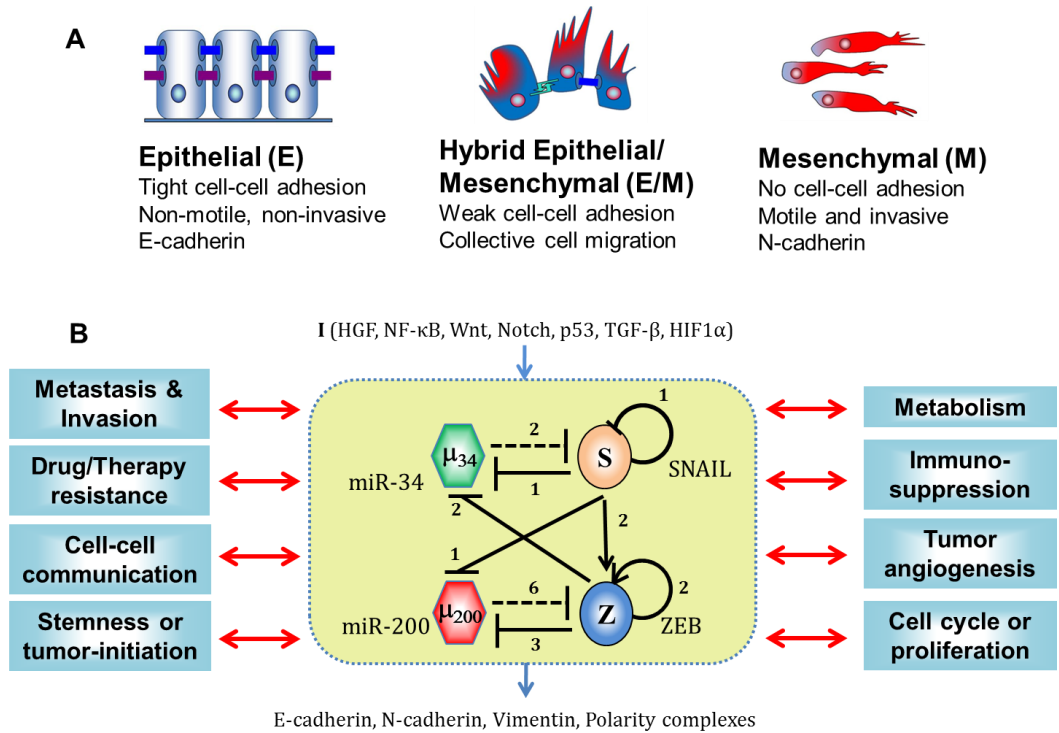

*Figure 1 EMT phenotypes and core EMT network. (A) Canonical morphological and functional characteristics of the three phenotypes – epithelial (E), hybrid epithelial/mesenchymal (E/M) and mesenchymal (M). (B) Core EMT regulatory network (shown in yellow box) consists of two interconnected mutually inhibitory feedback loops – (miR-34/SNAIL and miR-200/ZEB). Solid bars represent transcriptional inhibition, solid arrows represent transcriptional activation, and dotted lines denote miRNA-mediated regulation. Numbers mentioned alongside each regulation are the number of binding sites for that particular regulation, as experimentally determined or proposed. This core network receives inputs from a variety of signals (shown by I), modulates many cytoskeletal elements (E-cadherin, N-cadherin, Vimentin, and polarity complexes), and couples with many other cellular traits.*

**EMT decision-making: The operating principles**

Epithelial cells can undergo EMT under the influence of many signaling pathways such as TGFβ, EGF, HGF, Notch, FGF, Wnt, and IGF (17), and mechanical factors such as ECM density (18). These signals usually activate one of the EMT-inducing transcription factors (EMT-TFs) – TWIST1, SNAI1, SNAI2 (SLUG), ZEB1, ZEB2 (SIP1), Brachyury, Goosecoid, SIX1, and PRRX1 – that directly or indirectly repress E-cadherin, the hallmark of epithelial phenotype. Conversely, EMT can be inhibited by p53, MET-TFs such as GRHL2 and ELF5, and microRNA (miR) families such as miR-200 and miR-34 (19).

In many carcinomas, these signals converge on a core EMT regulatory network, also referred to as 'motor of cellular plasticity' owing to its coupling with many key cellular properties such as apoptosis, cell cycle, senescence and immunosuppression (20–24). This regulatory network is composed of two TF families – SNAIL and ZEB and two microRNA families – miR-200 and miR-34. The epithelial phenotype corresponds to high levels of miR-200 and miR-34, whereas the mesenchymal phenotype corresponds to high levels of ZEB and SNAIL. These components form two interlinked mutually inhibitory feedback loops – miR-34/SNAIL and miR-200/ZEB (25–27), such that EMT-inducing signals such as TGFβ, EGF, HGF, and Notch induce ZEB and SNAIL, and p53 activates miR-200 and miR-34 (Figure 1B).

*Mutually inhibitory feedback loops: a central motif of cell-fate decision*

Mutually inhibitory feedback loops between two fate-determining transcription factors (TFs) is one of the simplest gene circuits, and form a central motif in many cell-fate decisions. For instance, CDX2 and OCT4 control the fate of pluripotent embryonic stem cells - CDX2 induces trophoectoderm (TE) fate and OCT4 induces the opposite 'sister' fate - inner cell mass (ICM) (28). Similarly, cross-inhibitory TFs PU.1 and GATA.1 are situated at the branch point of erythroid and myeloid lineages in hematopoiesis. The mutual repression between the two TFs guarantees mutual exclusivity of the two identities (for instance, an erythroid cell cannot be a myeloid cell and *vice-versa*), and hence distinct cell identities (28). Therefore, a mutually inhibitory loop between two TFs A and B usually behave as binary or bistable switches allowing two distinct cell-fates – one corresponding to (high A, low B) expression and the other by (low A, high B), or in other words, (1, 0) and (0,1) states where "0" denotes relatively low expression, and "1" denotes high expression (29–31) (Figure 2).

There can be two important variations to this bistable behavior of a mutually inhibitory feedback loop. First, if mutual repression between the two TFs is not strong enough, both A and B are co-expressed at some intermediate level (½, ½) and the feedback loop does not give rise to two distinct cell fates (32,33). Second, if one or both TFs auto-activate themselves strongly in addition to strongly repressing the other TF, the circuit can allow for three distinct phenotypes – (1, 0), (0, 1) and (½, ½) – or (high A, low B), (low A, high B) and (medium A, medium B). The (½, ½) state can act as the 'poised' state of a progenitor cell that can differentiate to attain either of the two lineages – (1, 0) or (0, 1) (30,34) (Figure 2).

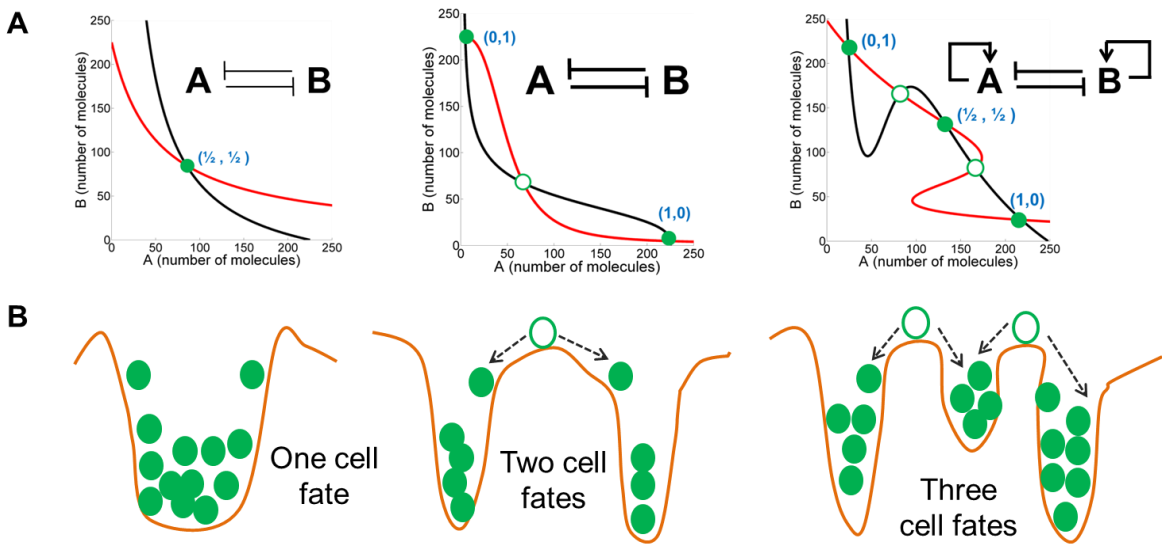

*Figure 2 Dynamic characteristics of mutually inhibitory feedback loops.* *(A) (left) Weak mutual inhibition between A and B allows monostability where the steady state has intermediate levels of both A and B; (middle) Strong mutual inhibition between A and B can drive one species to extremely low levels, and therefore bistability, such that the two steady states are – (high A, low B) or (1,0), and (low A, high B) or (0,1); (right) Intermediate mutual inhibition between A and B can strong self-activation of both A and B can enable the system to be tristable, such that the three steady states are - (high A, low B) or (1,0), and (low A, high B) or (0,1), and (medium A, medium B) or (½, ½). Red and black curves describe nullclines for A and B, and their intersections are the steady states. Green filled circles represent stable steady states, and green hollow circles show unstable steady states. The thickness of lines representing mutual inhibition between A and B, and self-activation of A and B, represent relative strength of those interactions. (B) Cartoons (corresponding to the circuit drawn in the same column) representing the potential energy of the system, where valleys represent stable steady states, and troughs denote unstable steady states.*

Importantly, distinct cell fates, as discussed above, are different than quantitative trait variation between two cells belonging to the same fate. For instance, CDX2 levels in two cells both belonging to trophoectoderm are most likely to be slightly different because of cellular stochasticity or non-genetic heterogeneity (35). However, neither of these cells spontaneously, or upon a small perturbation, transdifferentiate to adopt a different fate. Transdifferentiation often requires a large external signal such as overexpression of some cell-fate 'master regulator' transcription factors (36,37). This robust behavior of cell-fates reflects that they are 'stable steady states' of the underlying regulatory network, characterized by a particular range of values of all variables (expression levels) of all the elements in the system (genes, chromatin states, etc.) (28). Therefore, the existence of distinct cell fates in a cell population is often manifested as a multimodal distribution of some of these elements that can be captured in FACS experiments (Figure 3A). Conversely, if a system is monostable but exhibits quantitative trait variation as a function of external drivers, the FACS distribution will be roughly Gaussian (Figure 3B). Consequently, switching cell-fates usually entails a discontinuous jump in the expression levels of many genes, and can therefore be observed only in a multistable system, but not in stochastic variations within a monostable system (same cell fate). For instance, when a bipotent progenitor cell type differentiates to adopt any one of the two daughter lineages, the steady state corresponding to the bipotent progenitor disappears or loses its stability, and two new stable steady states emerge – each corresponding to a daughter lineage and each with a new and distinctive expression pattern (38) (Figure 3A).

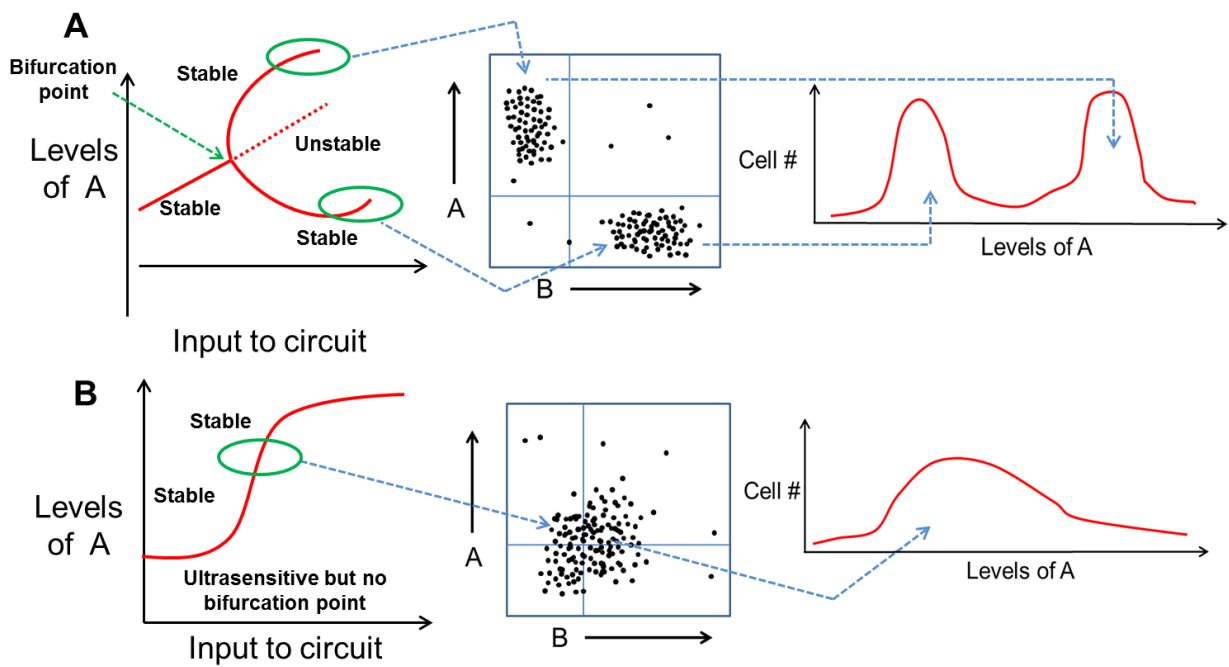

*Figure 3 Distinct cell fates vs. quantitative trait variation of same cell fate. (A) (left) Bifurcation diagram representing variation in levels of A as an input is applied to a mutually inhibitory circuit between A and B. At some threshold value of the input signal (marked by bifurcation point), the initial cell fate disappears and gives rise to two new stable steady state or cell fates. (middle) These two cell fates can be observed in a FACS experiment. (right) Most cells attain one of the two cell fates, and population distribution is bimodal with different range of values of A. (B) (left) Bifurcation diagram representing variation in levels of A as an input is applied to a mutually inhibitory circuit between A and B. The circuit responds in an ultrasensitive manner but no bifurcation of cell fates observed. (middle) FACS experiments show a population with continuously varying levels of A without any sharp boundaries, hence (right) the population distribution is unimodal and broadly Gaussian.*

These major qualitative as well as quantitative differences between distinct cell fates that emerge and then disappear as inputs are varied versus quantitative trait variation of a single fate have important implications. In a typical experiment, some specific attributes of the population are monitored as a control parameter is varied. Monostable systems exhibit no hysteresis and no multimodality in the population structure; they can however be ultrasensitive and thereby exhibit sharp thresholds in dose-response curve (Figure 4A). A system that exhibits multiple states with individual cells making fate decisions will in general exhibit hysteresis, will often exhibit multimodality, in addition to being able to exhibit sharp thresholds as the systems reaches points of bifurcation (Figure 4B). We shall argue below that the experimental data currently available for the EMT process strongly suggests an interpretation in terms of distinct cell fates, but this needs to be carefully addressed in more quantitative and carefully designed future experiments.

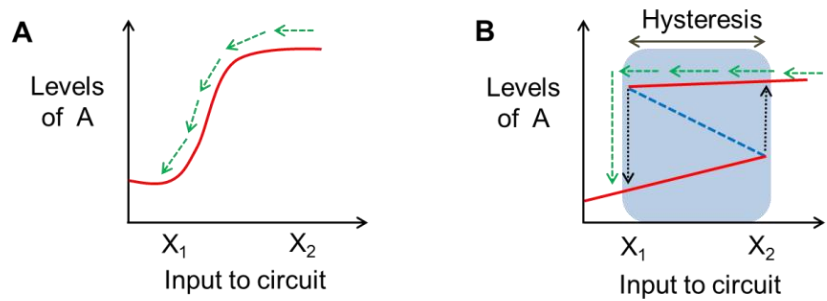

*Figure 4 Comparing the behavior of a phenotypic transition that has continuous state-space vs. the one with discrete state-space. (A) Behavior of a system with continuous state-space (infinite stable steady states) during a phenotypic transition as induced by an external input signal. Green arrows denote the response of the system when the input signal is removed. Solid red lines show stable states, and blue dotted lines show unstable steady states. (B) Behavior of a system with discrete state-space (finite number (n=2 here) of stable steady states (cell fates) to a varying input). Blue shaded region shows the range of hysteresis and bistability. Black dotted arrows mark the levels of input where the cell switches fate (or transitions from one stable steady state to another) – $X_1$, $X_2$. The table presented below compares the behavior of the two scenarios depicted in (A) and (B)*

### Why two mutually inhibitory loops in the core EMT network?

As mentioned above, the core EMT network comprises two mutually inhibitory loops – (miR-34/SNAIL) and (miR-200/ZEB). Two computational models of this network have proposed different functions for these two loops. Tian *et al.* (39) have proposed that both the loops – (miR-34/SNAIL) and (miR-200/ZEB) – function as bistable or binary switches that initiate and complete EMT respectively. They define the E phenotype as (high miR-200 and miR-34, low ZEB and SNAIL), M phenotype as (low miR-200 and miR-34, high ZEB and SNAIL) and partial EMT as (low miR-34 and ZEB, high SNAIL and miR-200). On the other hand, we propose that miR-34/SNAIL acts as a noise-buffering integrator of various EMT- and MET- inducing signals, preventing aberrant activation of EMT or MET due to transient signals, but not giving rise to any phenotypic transitions by itself. In other words, we argue that this subsystem would be monostable if it could be detached from any feedback from downstream effectors. Conversely, miR-200/ZEB, with input from SNAIL, behaves as a tristable or three-way switch allowing for the existence of three phenotypes – E (high miR-200, low ZEB), M (low miR-200, high ZEB) and E/M or partial EMT (medium miR-200, medium ZEB) (21,40).

Both existing models provide similar explanations for the E and M phenotypes, and can therefore be compared to experiments that focus on cells that undergo a complete EMT. Experiments showing that SNAIL can initiate repression of E-cadherin but ZEB is required for its complete inhibition (41), and that most genes repressed during EMT are inhibited by ZEB irrespective of the EMT-inducing signal (42), are consistent with either model; both approaches argue for ZEB activation to be necessary for a complete EMT (transition to a completely mesenchymal phenotype). Similarly, experiments showing that upon withdrawing the EMT-inducing signal, only the cells with low ZEB levels, but not high ZEB levels, revert to being epithelial immediately, indicate that ZEB activation marks a commitment point for cells to undergo an EMT (43) - a prediction both models make. Parenthetically, this lack of reversion is direct evidence in favor of the multistability picture (Figure 4B). Further, both the models predict that reverting EMT requires suppressing the EMT-inducing signal as well as ZEB, and that SNAIL knockdown does not suffice. Experiments validating these predictions, again, fail to discriminate between the two models (44–46).

However, experimental studies focusing on partial EMT can distinguish between the two models, and appear in our opinion to be more consistent with (medium miR-200, medium ZEB) definition of partial EMT rather than with (high miR-200, low ZEB). For example, studies in mammary morphogenesis – a canonical case of partial EMT – identified a transcription factor that can maintain the TEB (terminal end bud) cells in a partial EMT phenotype – OVOL, and knockdown of OVOL leads to complete EMT. Thus, OVOL acts as a "critical molecular brake on EMT" (47). OVOL is coupled with EMT core circuit in an intricate manner – it forms a mutually inhibitory switch with ZEB, inhibits miR-200 indirectly, and self-inhibits (48–51). Adding these interactions to our model, we showed that OVOL expands the range of parameters or physiological conditions for the existence of partial EMT or hybrid E/M phenotype (52), thereby explaining its role in maintaining the partial EMT phenotype. Similarly, co-expression of ZEB1 and E-cadherin in cells undergoing gastrulation (another example of partial EMT) (53) and, as discussed later in this review, the association of partial EMT with high tumor-initiating potential ('stemness'), are more likely to correspond to the (medium miR-200, medium ZEB) state structure for partial EMT.

The different results for partial EMT in the two models emerge from different modeling assumptions. The study by Tian *et al* (39) assumes simple universal forms for the various repressive interactions in the double-switch circuit. This assumption ignores key experimentally identified differences between the architecture of these two loops – two binding sites of miR-34 on SNAIL mRNA vs. six binding sites of miR-200 on ZEB mRNA(54,55), self-inhibition of SNAIL vs. (indirect) self-activation of ZEB (56,57), and finally the difference in transcriptional regulation vs. translational regulation by miRs (58–62). Importantly, the number of binding sites of a miR on an mRNA is crucial for determining the fold-change repression in protein expression, as shown by experiments that overexpression of miR-34 reduces SNAIL levels to 50% of the initial levels, but overexpression of miR-200 reduces ZEB levels to 10% (54,55). Further, the self-inhibition of SNAIL is critical to avoid any aberrant activation of EMT from transient activation of signals, and sets a sensitivity threshold for various EMT-inducing signals (41). Finally, the mechanisms of transcriptional regulation and miR-mediated sequestration and degradation of target mRNAs are distinct from each other and hence typically represented by different functional forms (30,58–62). Nevertheless, the manifestation of partial EMT state can be cell line-specific (63), because, for instance, not all cell lines might have same number of available miR-200 binding sites on ZEB mRNA, therefore, more quantitative measurement at the single-cell level is required to decipher which characterization of partial EMT holds in a particular context.

*Cellular heterogeneity during EMT*

Different levels of SNAIL enable different phenotypes and/or combinations thereof; for instance, low levels of SNAIL cannot induce an EMT, and very high levels can induce a complete EMT in almost the entire population (40). However, as observed in both physiological and pathological EMT contexts, the population can be highly heterogeneous, allowing for the emergence of distinct subpopulations of cells with different phenotypes. Cells in these distinct subpopulations may also interconvert their phenotypes due to intracellular stochastic fluctuations (16,64). Of course, different cell lines (or biological contexts) would be expected to have different ratios of these subpopulations (Figure 5A). Such cell-to-cell heterogeneity might have crucial functional consequences, especially in adaptive drug resistance, tumor dormancy and the heterogeneity in CTCs (35,65–68). These variations ride above purely genomic variations, which themselves can be quite extensive, given the compromised genome integrity in most cancers (1).

Recently, an important quantitative metric – an 'EMT score' – has been proposed to represent the overall proclivity of a cell line or primary tumor towards undergoing EMT, however, it largely ignores the cellular heterogeneity and possible clonal heterogeneity inherent to a particular cell line (69). Not surprisingly, these scores vary continuously. Given the evidence described above that the E and M states are truly different cell fates, the continuous variation argues in favor of additional stable intermediate states that occupy different positions on 'EMT axis' (63,69); otherwise, we would in general expect to see a sharper score variation. This is of course what we have already expected based on the circuit models and based on the analogy between pathological EMT and the physiological EMT examples of wound healing and branching morphogenesis. It remains to be investigated precisely how many stable intermediate states are present *en route* EMT and whether this inference is proven correct by individual cell studies. Also, it must be noted that unlike developmental EMT, pathological EMT might not necessarily involve a real lineage-switching of cells in an epithelial lineage to a mesenchymal one (70).

Another related important question that needs to answered is that how morphologically stable is (are) the intermediate state(s) of EMT. Partial EMT has been usually labelled as a 'metastable' state (8), indicating that it is less stable than pure E or pure M ones. However, recent experimental studies have identified that some epigenetic changes (71) as well as some 'phenotypic stability factors' such as OVOL (72) can fine-tune the transitions into and from partial EMT. Cells expressing endogenous levels of OVOL can maintain their partial EMT phenotype, knockdown of OVOL leads to complete EMT and overexpression of OVOL induces the reversal of EMT – a MET (47,48). These experimental findings can be unified via our theoretical framework by coupling OVOL to the core EMT network, where we show that OVOL can both act as a "critical molecular brake on EMT" preventing the cells "that have gained partial plasticity" to undergo a complete EMT, and a driver of MET when overexpressed (47,52) (Figure 5B). Our work on

OVOL serves as an example of how our theoretical framework for the core EMT network renders itself to analyzing the role of other regulatory players in epithelial plasticity(52).

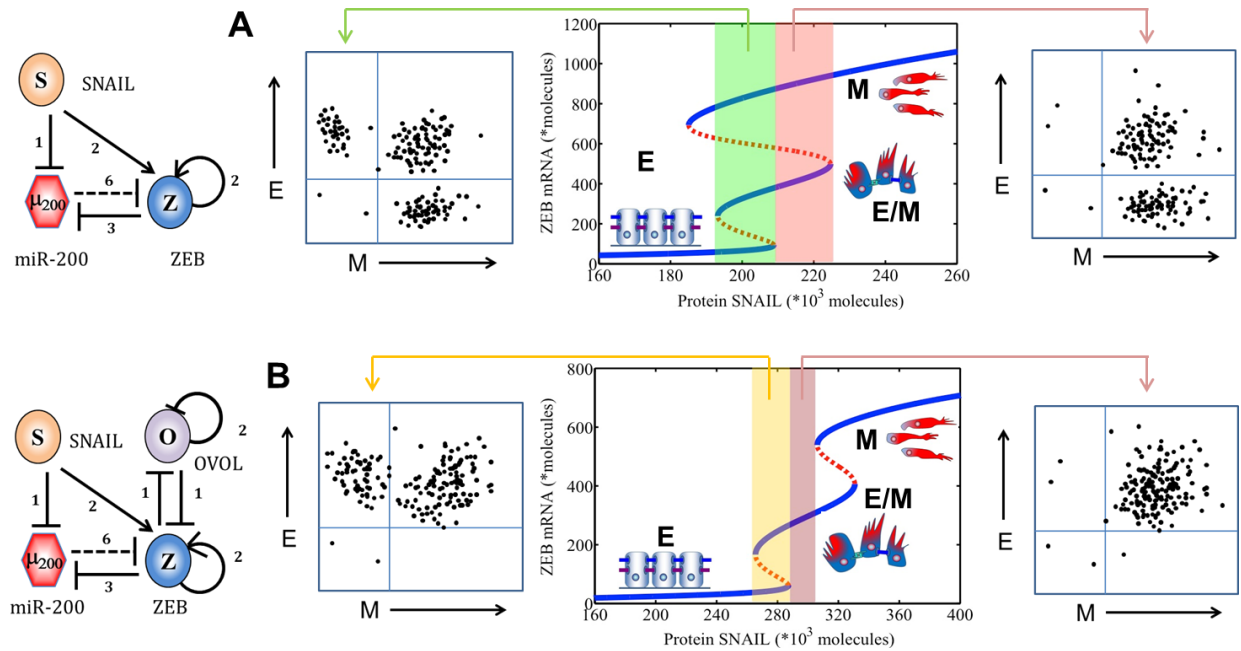

*Figure 5 Population distribution or multimodality in EMT response. (A) (middle) Bifurcation of ZEB mRNA levels in response to protein SNAIL (EMT-inducing signal) for the miR-200/ZEB/SNAIL circuit (shown at extreme left). For a certain range of SNAIL values (marked by green rectangle), cells can attain any of the three phenotypes – E, M and E/M, giving rise to a trimodal population distribution as shown in FACS figure (left). For a different range of SNAIL values (marked by orange rectangle), cells can adopt either E/M or M phenotype, and be distributed in a bimodal manner in FACS figure (right). (B) Bifurcation of ZEB mRNA levels in response to protein SNAIL (EMT-inducing signal) for the miR-200/ZEB/SNAIL/OVOL circuit (shown at extreme left). For a certain range of SNAIL values (marked by yellow rectangle), cells can adopt either E/M or E phenotype, and be distributed in a bimodal manner (FACS figure, left); and for a different range (marked by red dotted rectangle) all cells are likely to be in E/M phenotype as shown in FACS figure (right). Importantly, as compared to the behavior of miR-200/ZEB/ SNAIL circuit, miR-200/ZEB/SNAIL/OVOL circuit allows the existence of new phases (combinations of phenotypes) such as {E/M} and {E, E/M}, and precludes the existence of phases {E, E/M, M}.*

**EMT effects on cellular shape and behavior**

Cells that become motile as a result of (complete) EMT appear to come in two distinct shapes and concomitant behaviors, namely mesenchymal and amoeboid (73). Note that there is no guarantee that cells described as M from the genetic network perspective always have mesenchymal shapes. Cells labeled as mesenchymal are spindle-shaped, have lamellopodia and/or filopodia on their leading edge, adhere strongly to the ECM, and act as 'path generators' by secreting Matrix Metallo-proteinases (MMPs). Conversely, amoeboid cells are round-shaped, often have blebby structures, have low adhesion to ECM, and show a higher shape plasticity that helps them squeeze through the gaps in ECM and act as 'path finders' (74,75). Further, cells can adopt a shape representing both amoeboid and mesenchymal traits (hybrid A/M) such as cells with both lamellopodia and blebs (76). In cancer, there is a rich plasticity that allows cells to adopt functional behaviors depending on external signals, phenotypic choices, and of course genetic changes – such as switching between amoeboid and mesenchymal morphologies – a Mesenchymal to Amoeboid Transition (MAT) and its reverse – AMT, and direct bidirectional switching between hybrid E/M and A phenotypes – a Collective to Amoeboid Transition (CAT) and its reverse – ACT (77–80). Presumably, this plasticity enables them to adapt to different environments encountered during metastasis, and is therefore critical for tumor dissemination (78) (Figure 6A).

Elucidating the principles of this plasticity requires investigating the coupling between the core EMT circuits and the downstream effectors actually responsible for actualization of motility biophysics. One key piece is the mutually repressing feedback loop between the two GTPases – RhoA and Rac1 – that promote their own GTP loading and inhibit that of the other. Activation of RhoA increases actomyosin contractility resulting in membrane blebbing and facilitating a rounded amoeboid phenotype. Conversely, activation of Rac1 results in focal adhesions and actin polymerization, leading to formation of lamellopodia, enabling a front-back polarized spindle-shaped mesenchymal cell (73,81). Importantly, these two GTPases play crucial roles during EMT in converting apico-basal polarity typical of cells in epithelial layers to front-back polarity needed for motion. For instance, Rac1 activation at the leading edge stimulates PI3K that leads to indirect self-activation of Rac1, therefore setting up a positive feedback loop for cytoskeletal reorganization necessary for cells to gain directional migration abilities. Similarly, RhoA promotes actin stress fiber formation and prevents formation of the polarity complexes PAR at the rear end of the cell (82). Quite reasonably, the epithelial gatekeepers miR-34 and miR-200 inhibit the translation of RhoA and Rac1 (83–85) (Figure 6B). However, the resultant dynamics of this complex interplay remains a challenge for the future. In particular, a comprehensive understanding of cell shape dynamics and its coupling to EMT will require when epithelial cells attain partial EMT as well as during their transition to these solitary migration phenotypes requires integrating live-cell imaging with a multi-compartment spatiotemporal model capturing the spatial segregation of the GTPases (86–88).

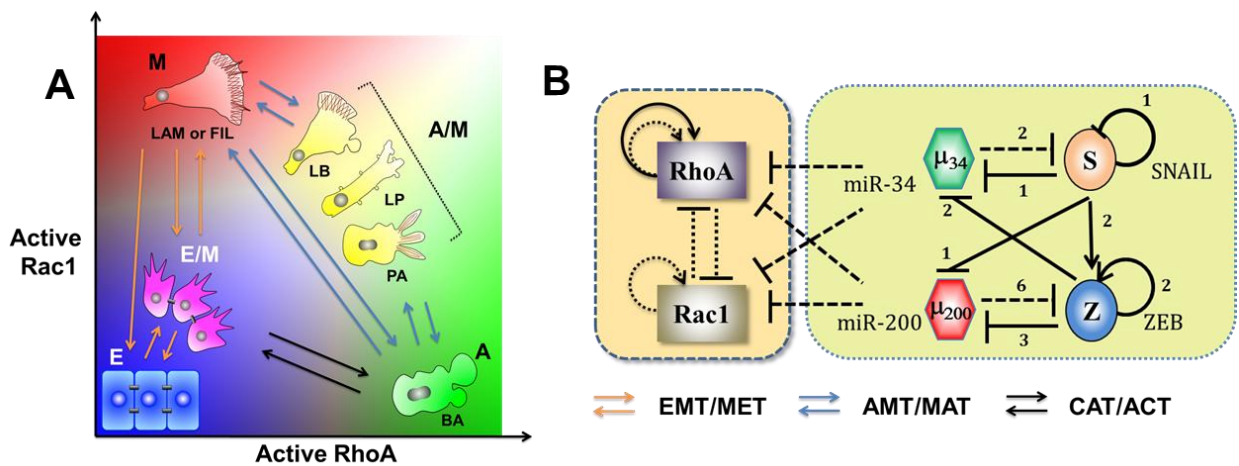

*Figure 6 Landscape of cellular shape plasticity during carcinoma metastasis.* *(A) Cartoon representation of different cell shapes/phenotypes with their respective places on the two-dimensional space of levels of active RhoA (RhoA-GTP) and active Rac1 (Rac1-GTP). As miR-34 and miR-200 inhibit both RhoA and Rac1, both epithelial and hybrid E/M phenotypes have low levels of active forms of RhoA and Rac1. The (High RhoA-GTP, low Rac1-GTP) profile associates with amoeboid (A) morphology with blebs (blebby amoeboid (BA)), whereas (low RhoA-GTP, high Rac1-GTP) associates with mesenchymal (M) shape – cells with lamellopodia or filopodia (LAM or FIL). Cells with (high RhoA-GTP, high Rac1-GTP) adopt a hybrid A/M morphology that can be manifested in multiple ways- lamellipoida with blebs (LB), lobopodia (LP) and pseudopodal amoeboid (PA). Transitions among E, E/M and M phenotypes (EMT/MET) are represented by orange arrows, those between amoeboid and mesenchymal morphologies – A, A/M and M – are denoted by blue arrows, and transitions between E/M and A phenotypes – CAT/ACT – are denoted by black arrows. (B) Circuits showing the coupling of core EMT circuit with RhoA and*

*Rac1 – the two GTPases that are critical in regulating cell shape. They inhibit the GTP loading (switching from inactive GDP-bound state to active GTP-bound state) of each other and promote that of themselves (shown by dotted lines). Also, RhoA can activate itself indirectly on a transcription level (solid black lines) (See Ref. (73) and references therein). The microRNAs miR-34 and miR-200 inhibit the translation of RhoA and Rac1.*

**Partial EMT allows collective migration during development**

The partial EMT phenotype – (medium miR-200, medium ZEB) – has been studied extensively in embryonic development and wound healing (6,89,90). A canonical process showing the role of partial EMT in development is the branching morphogenesis of the trachea and mammary gland – a mechanism that enables the repeated splitting of a tubular epithelial structure to generate a ductal tree. During branching morphogenesis, the tip cells located at the cap of the Terminal End Buds (TEBs) of the growing tubule maintain cell-cell adhesion with neighbors and transiently display mesenchymal features such as loss of apico-basal polarity and increased motility in response to extracellular signals such as FGF (6,7). These collectively migrating cells express P-cadherin, a proposed marker for partial EMT (9,91), and form finger-like projections and maintain their partial EMT phenotype (i.e. do not proceed to a complete EMT) possibly due to the action of a "critical molecular brake on EMT" – the transcription factor OVOL – whose knockdown leads to solitary and impaired migration (47). Similar to TEB migration, during sprouting angiogenesis 'tip' endothelial cells display a partial Endothelial to Mesenchymal Transition (pEndMT) transition and lead the collective migration of a train of 'stalk' cells (92). Further, in wound healing, immature basal keratinocytes at the wound edge partially remodel their basement membrane and migrate collectively in a 'metastable' partial EMT phenotype, and finally, revert to being epithelial or, in other words, undergo re-epithelialization to close the wound (7,8,93,94). Collective migration in most partial EMT cases is mediated by SLUG (SNAIL2) (92,95–99). Such collective migration has multiple advantages – it obviates the need for all cells to detect external signals for migration, allows coupling of mechanical forces among the cells, and provides them with maximum plasticity to be able to switch to being epithelial or mesenchymal (complete EMT) phenotypes (6). These advantages can be utilized by carcinomas during invasion and intravasation of multicellular strands (100).

**Partial EMT enables migration of CTC clusters during metastasis**

Recent studies have highlighted the crucial significance of partial EMT in cancer metastasis. Cells co-expressing various epithelial and mesenchymal markers are present in primary breast and ovarian cancer (13,101), in multiple cell lines belonging to ovarian, lung and renal cell carcinoma (63,102,103), as well as in mouse models of pancreatic ductal adenocarcinoma (PDAC) and prostate cancer (104,105). Importantly, among breast cancer subtypes, the ones with poor clinical outcomes – triple-negative (TNBC) and basal-like (BLBC) – are most enriched for such biphenotypic cells, indicating a strong association between aggressiveness and E/M phenotype (13,106,107). Further, co-expression of mesenchymal marker vimentin and epithelial/luminal markers cytokeratins 8 and 18, rather than the expression of vimentin alone, correlates with increased invasive and metastatic potential and poor survival and is often observed in many aggressive tumors such as BLBC and melanomas (108–112). Besides, a gene signature consisting of both epithelial and mesenchymal genes predicts poor outcomes independent of breast cancer subtype (16), suggesting that association of partial EMT phenotype with aggressiveness can be context-independent.

Cells co-expressing E and M markers can also be present in the bloodstream of breast, lung, colon, and prostate cancer patients as clusters of CTCs that contain a median level of three cells per cluster (10–

13,113). These clusters, also referred to as 'microemboli', can be apoptosis-resistant, are more likely to be trapped in narrow blood vessels for extravasation, and often correlate with poor prognosis in patients (11,14,15,114). Although these clusters constitute only 3% of total CTC 'events' observed (97% being individually migrating CTCs), they contribute 50% of the total metastases, reflecting their increased metastatic propensity (15). Further, they can be found in the bloodstream of patients with COPD (chronic obstructive pulmonary disease) around three years before a lung nodule can be detected, and therefore they might be useful to identify patients at a greater risk of developing lung cancer (115). Importantly, the heightened metastatic potential of such clusters as compared to that of the same numbers of individual cells was recognized (116,117) even before EMT was characterized as a metastasis mechanism (118).

However, CTC clusters need not necessarily contain only the hybrid E/M cells, and a comprehensive understanding of other cell types that might be present in these clusters is necessary for advancing the clinical application of CTCs analysis as a 'liquid biopsy' (14,119). There may be admixtures of E and M cells in a single cluster. Also, representing the cellular heterogeneity of the primary tumor, some of these clusters may contain leukocytes as well as platelets and megakaryocytes (13,120), therefore true "seeds" of metastasis (metastasis initiating cells) within these CTCs must be identified carefully using functional assays such as xenotransplantation in immunodeficient mice. Initial attempts in this direction have elucidated that the cancer cells in a hybrid E/M phenotype (identified by $SNAIL^+$ $E\text{-}cad^+$ in colon, and by $EPCAM^{lo}$ $MET^{high}$ in breast cancer) can more efficiently act as seeds of metastasis (121,122), hence establishing a clinical and prognostic relevance of the cells in the hybrid E/M phenotype. Nonetheless, not all cells in a hybrid E/M phenotype might be capable of initiating a tumor *in vivo* (123).

**Partial EMT, but not necessarily complete EMT, associates with stemness**

A subpopulation of cancer cells that seed metastasis or in other words, have self-renewal as well as clonal tumor initiation ability along with long-term clonal repopulation potential are referred to as Cancer Stem Cells (CSCs). These cells with stem-cell properties ('stemness') can evade cell death and cancer therapeutics, and may stay dormant for long periods of time (124). However, in the context of cancer, 'stemness' is not a fixed inherent trait of a few privileged cells, rather CSCs and non-CSCs can interconvert among themselves, and this plasticity or dynamic equilibrium drives tumor growth as well as invasion (125–128). Functional assays of isolating CSCs include mammosphere-formation *in vitro* and limiting dilution assays of tumor-initiating potential *in vivo* in NOD/SCID mice. In other words, CSCs are usually characterized by high evolvability (capacity to give rise to heritable phenotypic variation) (129).

Under some conditions, cells undergoing a full EMT have been shown to be highly likely to gain 'stemness' and behave operationally as Cancer Stem Cells (CSCs). This EMT-stemness coupling was first reported for immortalized human mammary epithelial cells (130,131). Similar findings in many carcinomas such as pancreatic, hepatocellular and colorectal have strengthened this notion (132). Therefore, aberrant activation of EMT can serve at least two functions – (a) increases the invasion ability to reach distant organs for metastasis, and (b) enhances tumor-initiating properties of the cells that reach the metastatic sites (133). However, this notion of a full EMT coupled with stemness has been challenged by studies showing that repression of EMT is required for effective tumor-initiation (134–137) and that reprogramming often involves MET (138,139).

A few recent studies attempt to resolve this contradiction by suggesting that instead of the cells in pure epithelial (E) or pure mesenchymal (M) states, cells in hybrid E/M or partial EMT state are most likely to gain stemness (16,105,123,140). Grosse-Wilde *et al.* show that co-expression of E and M genes in the very same cell promotes mammosphere formation and stemness, independent of the breast cancer subtype (16). Further, Strauss *et al.* showed that some cells in hybrid E/M phenotype in primary ovarian cultures and tumors *in situ* can be multipotent, express markers of other lineages, and drive tumor growth *in vivo* by giving rise to another E/M subset as well as completely differentiated epithelial cells (123). Ruscetti *et al.* isolated hybrid E/M cells *in vivo* in a prostate cancer mouse model and demonstrated their comparable or even higher sphere formation and tumor-initiating potential as compared to completely mesenchymal cells (105). Also, our study that mathematically models the stemness-decision circuit (LIN28/let-7) with inputs from miR-200 and NF-κB suggests that especially at high levels of NF-κB, hybrid E/M state is more likely to gain stemness than complete EMT (140). These studies propose that cells undergoing partial EMT, but not necessarily complete EMT, can gain stemness, or in other words, the 'stemness window' lies somewhere close to midway on the 'EMT axis' (Figure 7) (133); and are consistent with experiments showing that more than 80% CTCs in men with castration resistant prostate cancer (CRPC) and over 75% of CTCs in women with metastatic breast cancer co-express epithelial markers cytokeratins (CK), mesenchymal markers N-cadherin, and stem cell markers (12).

This association of hybrid E/M phenotype with stemness is not specific to tumor progression, but has also been reported in physiological EMT examples where adult hepatic stem/progenitor cells (HSCs) co-express epithelial and mesenchymal genes, and give rise to both epithelial and mesenchymal lineages in the liver (141–144). Similar to HSCs, adult renal progenitors are in a 'metastable' hybrid E/M state upon tissue injury and mediate renal repair and regeneration (145). Collectively, these studies present strong evidence for the emerging notion that CTCs in a semi–mesenchymal phenotype, rather than those 'frozen' or locked in a full EMT phenotype, have the highest plasticity to switch between proliferative and invasive modes, are capable of completing the invasion-metastasis cascade, and should therefore be regarded as CSCs (119,146–149).

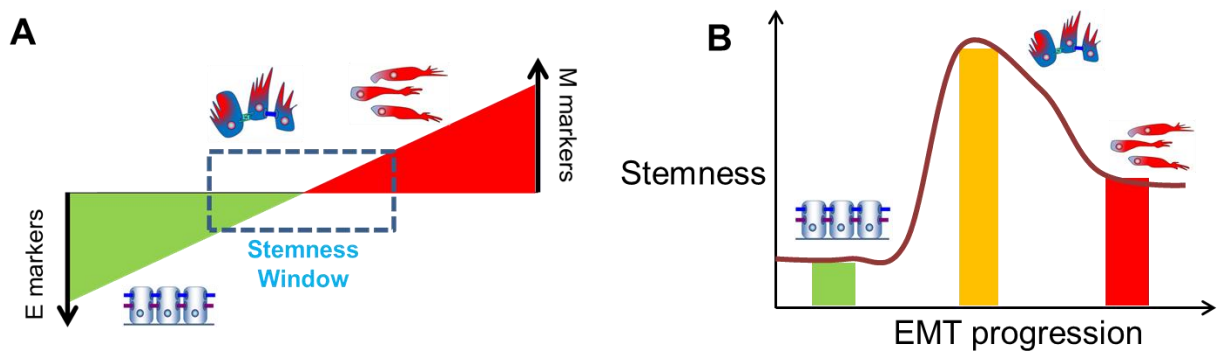

*Figure 7 Association of partial EMT with stemness.* (**A**) *'Stemness window' model proposed by Ombrato and Malanchi (133) where stemness is maintained within a window between a fully differentiated epithelial cell and a fully de-differentiated mesenchymal cell (Figure adapted from Ref. (133),cartoons included).* (**B**) *Stemness or tumor-initiating potential of E, E/M and M phenotype, or variation of stemness during EMT progression (brown line).*

An alternative hypothesis that attempts to resolve the connection between EMT, MET and stemness proposes that CSCs come in two distinct states – 'epithelial-like' and 'mesenchymal-like' (150,151). Importantly, these studies show that most epithelial-like CSCs (identified by ALDH[+] by Liu *et al.* (150) and by CD44[high] EPCAM[high] by Biddle *et al.* (151)) can give rise to both epithelial-like and mesenchymal-like populations and hence bilineage colonies *in vitro*. However, this plasticity was significantly impaired in the mesenchymal-like CSCs (identified by CD44$^+$/CD24$^-$ by Liu *et al.* (150) and by CD44$^{high}$ EPCAM$^{low}$ by Biddle *et al.* (151)), thereby contributing to the notion that cells locked in a complete EMT phenotype significantly lose their plasticity. Neither of these studies considered the possibility that there might exist hybrid E/M states, but did show that not all CSC's have to have the same EMT properties. At present, the exact mapping between EMT and stemness appears to be complex and context-dependent (4,152), but with evidence suggesting that the major target to avoid tumor relapse and metastasis might be the CTCs in a hybrid E/M phenotype which have same degree of stemness (153). It, therefore, becomes essential to find a set of markers that would enable such cells to be identified.

**Proposed CTC markers for partial EMT cells**

Identifying a robust set of markers to isolate CTCs in a hybrid E/M phenotype remains an open question. CTCs with an epithelial phenotype can be identified via cell surface markers such as EPCAM (epithelial cell adhesion molecule) or cytoskeletal markers such as CK8, CK18, and CK19. However, establishing similar markers for CTCs with at least a partially mesenchymal phenotype has been challenging because vimentin is expressed in most normal blood cells as well (154). A potential signature for identifying the 'stem-like' hybrid E/M CTCs may be CD24$^+$CD44$^+$, the expression pattern for pancreatic and gastric CSCs (155,156), because CD24 is a canonical epithelial marker, and CD44 is a mesenchymal stem cell one (157,158), and CD24$^+$CD44$^+$ expression pattern overlaps with high levels of P-cadherin, another proposed marker of partial EMT phenotype (91). Recent studies have highlighted that CD24$^+$CD44$^+$ cells can have up to ten times higher mammosphere initiating capacity, and can form more aggressive tumors than CD44$^+$/CD24$^-$ cells (16,65) that have been traditionally considered to be CSCs (159). CD24$^+$CD44$^+$ cells are present in multiple cell lines belonging to the luminal and basal-like subtypes, and their population is enriched significantly upon exposure to an acute cytotoxic shock, suggesting that they represent a drug-tolerant subpopulation that can repopulate a tumor (65). Collectively, these studies show that CD44$^+$/CD24$^-$ expression does not necessarily correlates with tumorigenicity (160), and consolidate the mounting evidence that cells with a biphenotypic E/M expression tend to have high tumorigenicity in mice (156,161–163).

While establishing a robust set of markers such as CD24$^+$CD44$^+$ for detecting the E/M cells in CTCs, at least two cautionary steps must be taken. First, a more quantified characterization of the presence of markers is required to identify the intermediate state(s) of EMT. For instance, CD24$^{neg}$ cells must be segregated from CD24$^{lo}$ cells, as they mark different lineages in mouse mammary gland, have dissimilar tumorigenic potential and respond differently to gamma-secretase inhibitors (GSI) due to their distinct gene expression profiles (158,161). Second, the clusters of CTCs need to be isolated and investigated for different cell types present in them. Owing to their residual cell-cell adhesion, the CTCs in a hybrid E/M phenotype are likely to attach to cancer cells and/or stromal cells to form CTC clusters. Therefore, isolating CTC clusters should have two major advantages – (a) capturing hybrid E/M cells that are not necessarily present on the surface of the cluster, and (b) revealing novel insights into the cooperation of cancer cells and/or cancer cells and stromal cells present in the same cluster. Such cooperation is expected to recapitulate the tumor-stroma ecology seen in primary tumors and metastasis, where some stromal cells can be 'activated' by cancer through cytokines to provide metabolic synergy and signals for survival and maintaining stemness (164–167). 'Activated' stromal cells can also be carried along as the 'soil' by the accompanying 'seed' metastatic cells to gain early growth advantage during colonization (168).

**Role of cell-cell communication in maintaining partial EMT**

Cell-cell communication among cancer cells and/or between cancer cells and stromal cells (fibroblasts, immune cells, endothelial cells etc.) can have a significant influence on phenotypic plasticity (EMT/MET), CSC self-renewal, and a dynamic equilibrium between CSCs and non-CSCs. Spatial heterogeneity in the tumor can lead to spatial variations of secreted factors, cell types that are in direct contact, ECM density etc. each of which can affect the 'EMT score' of individual cancer cells in the tissue (18,169). A key signaling pathway that is involved in multiple aspects of this cross-talk both via cell-cell contact and via soluble factors is Notch signaling. Notch signaling can induce EMT and maintain stemness (170); however, our understanding of the different roles of the two sub-families of ligands (Delta and Jagged) of Notch signaling is only recent (171–173), and incomplete in the context of EMT and/or CSCs.

Notwithstanding this lack of knowledge, Jagged1 is emerging as a potential therapeutic target for its roles in maintaining and increasing CSCs, inhibiting apoptosis, inducing angiogenesis, and affecting the immune cells (171). It can both be secreted by endothelial cells as well as present on the membrane of stromal and cancer cells and can activate Notch signaling in cancer cells to increase the CSC population (174,175). Further, it is implicated in colonization where it is present on the surface of breast cancer cells and can activate Notch signaling in the bone (176), correlates with poor survival outcome, is overexpressed in CSCs, and has much higher levels in the more aggressive forms of breast cancer such as triple negative (TNBC) and basal-like (BLBC) than in its luminal subtypes (171,177).

Importantly, Notch-Jagged (N-J) communication might be the preferred mode of tumor-stroma signaling than Notch-Delta (N-D) signaling due to its multiple potential synergistic effects in the tumor ecology. Two cells interacting via N-D signaling usually adopt distinct fates – one cell behaves as Sender (high ligand (Delta), low receptor (Notch)) and the other as Receiver (low ligand (Delta), high receptor (Notch)), therefore allowing only one-directional signaling and 'salt-and-pepper' cell-fate patterns (178) (Figure 8B). Conversely, the two cells interacting via N-J signaling can adopt similar fates– hybrid Sender/Receiver (medium ligand (Jagged), medium receptor (Notch)) that enables bidirectional communication between them (172,173), and allows lateral induction, i.e. a cell induces its neighbor to adopt the same fate as that of its own (179–181) (Figure 8C). Due to this lateral induction mechanism observed in N-J signaling, a cluster of E/M cells interacting via N-J signaling might mutually stabilize their 'metastable' phenotype and consequently maintain high 'stemness' (172). This notion is supported by the involvement of Notch signaling in wound healing (182). N-J signaling in collectively moving cells can induce or maintain similar fates as that of the neighboring cells, thereby coordinating wound healing, but excessive N-D signaling might impair it. Importantly, if partial EMT is defined as (high miR-200, low ZEB) rather than (medium miR-200, medium ZEB), collectively moving cells with active Notch signaling are likely to have suppressed N-J signaling almost completely (because miR-200 inhibits Jagged1 strongly(183)) and therefore might diversify their fates via N-D signaling, a phenomenon that would impair wound healing (40) (Figure 8A,D). A tantalizing possibility nevertheless, it remains to be tested both experimentally and via a theoretical model of the coupled core EMT circuit and Notch-Delta-Jagged signaling via interactions such as miR-200 inhibits Jagged (183), miR-34 inhibits Delta and Notch (184,185), and NICD activates SNAIL(186,187).

Notch-Jagged signaling can also mediate tumor-stroma interaction via regulating the secretion of many cytokines that can enslave or 'activate' stromal cells (124). For instance, IL-6 secreted by cancer cells drives the activation of normal fibroblasts towards becoming Cancer-Associated Fibroblasts (CAF) that in turn elicit an EMT response in cancer cells and increases the CSC population (164). Further, IL-6 can also promote the generation of Tumor-Associated Macrophages (TAM) that support tumor metastasis (188). Many inflammatory cytokines such as IFN-γ and IL-6 can also increase production of Jagged and/or decrease that of Delta (177,189,190) hence possibly forming a positive feedback loop that rakes up Notch-Jagged signaling and mediates chronic inflammation, a hallmark of cancer, in the stroma (1,191).

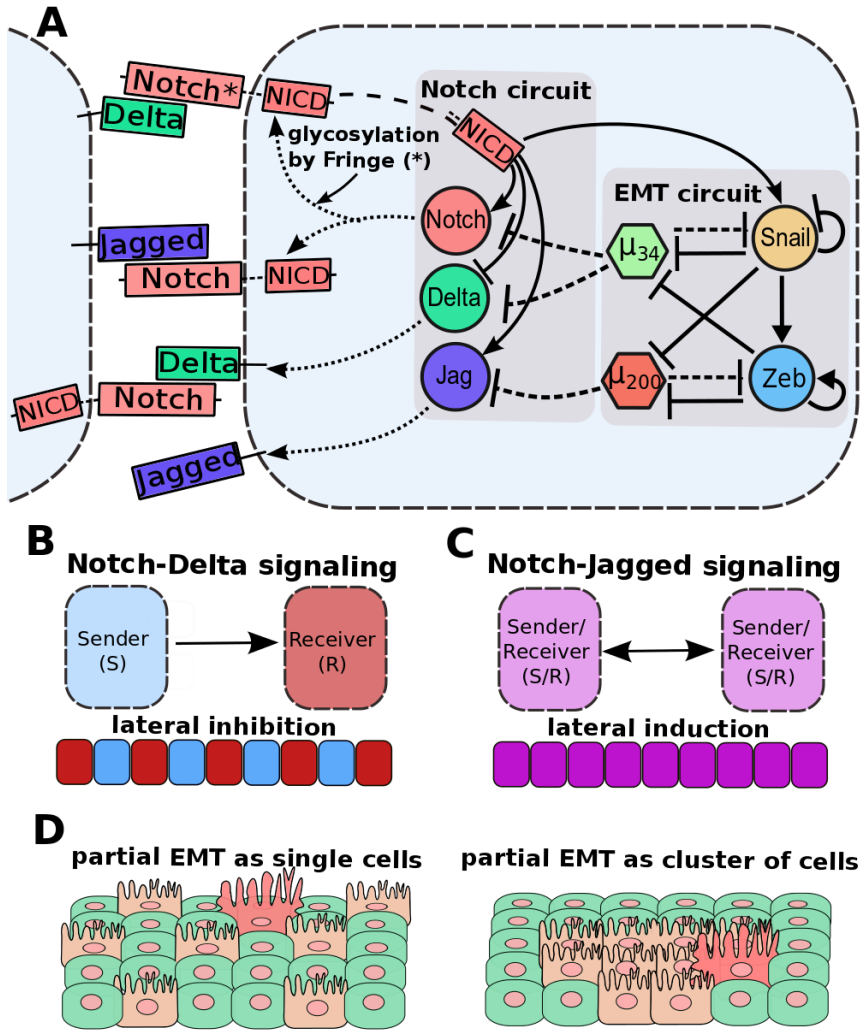

*Figure 8 Cell-cell communication and partial EMT. (A) Coupling of EMT circuit with Notch circuit. Notch pathway, when activated by Jagged or Delta, belonging to neighboring cell, can activate Jagged and Notch, but inhibit Delta. EMT circuit couples with Notch circuit in many ways – miR-200 inhibits Jagged1, miR-34 inhibits both Notch and Delta, and NICD can activate SNAIL to drive EMT. (B) Notch-Delta signaling between two cells induces opposite fates in them – one cell behaves as a Sender (low Delta, high Notch) and the other a Receiver (high Notch, low Delta). Due to this lateral inhibition, it can promote 'salt-and-pepper' based patterns. (C) Notch-Jagged signaling between two cells induces similar fates in them – lateral induction – and thus lead to patterns with all cells with the same fate. (D) (left) Cells in a partial EMT and interacting via N-D signaling might not be able to spatially close to each other, because N-D signaling inhibits two neighbors to adopt the same fate. (right) Cells in a partial EMT and interacting via N-J signaling can mutually stabilize the E/M phenotype and stay together as a cluster.*

Therefore, cancer cells can be considered as 'ecological engineers' taking advantage of its niche in multiple ways, such as metabolic synergy and gaining CSC and migration traits (165–167). The outcome of such dynamic cross-talk can be best understood using an integrated computational and experimental approach – reconstructing the ecological dynamics of cancer via co-culture experiments, and building a multi-scale model combining intracellular signaling with population level spatial models. Such an understanding might provide valuable insights into therapies targeted at managing the stroma, as well as combinatorial therapies targeting both the cancer and stroma to avoid tumor relapse (165,192,193).

**Interplay between partial EMT and drug resistance**

EMT has been posited to be involved in drug resistance (194–196), however, characterizing cell lines based on EMT scores indicates that this correlation might not be universally applicable in all carcinomas (69). Importantly, most studies connecting EMT to drug resistance has viewed EMT as an 'all-or-none' process (194–196), leaving little scope for assessing the possible drug resistance in partial EMT phenotype, and comparing it with that corresponding to a complete EMT phenotype. Because CSCs have been reported to be primarily responsible for drug resistance (197), the association of hybrid E/M phenotype with stemness (16,105,123,140) proposes that a hybrid E/M or partial EMT phenotype can also be the phenotype maximally correlated with drug-resistance.

The partial EMT or 'EMT-like' phenotype can associate with both *de novo* and adaptive drug resistance. Among various breast cancer subtypes, the triple-negative breast cancer (TNBC) contains the maximum number of hybrid E/M cells in the primary tumor (13), and exhibits *de novo* resistance to current standard therapies such as anthracyclines and taxanes (198). There is also a strong relationship to adaptive therapy. Significantly, a paradigm that emerges from many recent studies is that cancer cells that become resistant to many therapeutic assaults often undergo partial EMT. Development of tamoxifen-insensitivity in MCF7 breast cancer cells and that of trastuzumab-resistance in HER2-overexpressing breast cancer cells is usually accompanied by a partial EMT (199,200). Further, the radiation-resistant colorectal cancer cells generate cellular progeny with an 'EMT-like' phenotype (201), and exposure to taxanes induce a phenotypic transition to a chemotherapy-tolerant state (CD44+ CD24+) in multiple cell lines belonging to both basal-like and luminal subtypes (65). It must be noted that CD44+ CD24+ expression pattern is proposed is what we proposed above to be a hallmark of hybrid E/M cells in many cancer subtypes (16).

The underlying signaling pathways and molecular mechanisms of this interplay between partial EMT and drug resistance remain largely elusive. It is not even clear what aspects of drug resistance are phenotypic in character (reduction of growth rate, upregulation of pumps etc.) and what depend on actual genetic changes and whether these are coupled via regulation of genomic instability. At the signaling level, key intermediary pathways involve Notch-Jagged signaling (171), that as discussed above, can play a key role in shepherding the epithelial-mesenchymal plasticity by stabilizing a 'metastable' partial EMT phenotype.

**Completing the loop: Inflammation, Notch-Jagged signaling, Partial EMT and stemness**

A key difference in partial EMT during wound healing and that during tumor progression is that during wound healing, cells often re-epithelialize after closing the wound, thereby limiting their plasticity, but during tumor progression, this ubiquitous plasticity spearheads aggressive tumor progression (93,202). Further, wound healing often elicits an acute inflammatory response that is resolved later (93); however, during cancer, 'the wounds that do not heal'(203), inflammatory response is chronic and is a hallmark of cancer (1,191). Therefore, inflammation can regulate the timespan of heightened epithelial plasticity and more specifically, the timespan over which a hybrid E/M or partial EMT phenotype can be maintained. Such a 'stabilizing' effect of inflammation on the 'metastable' partial EMT phenotype can be mediated largely by Notch-Jagged signaling, because many inflammatory factors such as TNF-α, IFN-γ, and IL-6 can increase the production of Jagged and/or decrease that of Delta (177,189,190), thereby promoting Notch-Jagged signaling that can maintain a cluster of cells in a partial EMT phenotype. Consistently, the breast cancer subtype that has maximum number of cells co-expressing E and M genes among all breast cancer subtypes (13) – TNBC – has elevated levels of Jagged1 as well as NF-kB (204).

Inflammatory stress conditions in tissues are also created by both chemotherapy and radiation by activating NF-κB, the central link between inflammation, tumor progression, and radiation-resistance (205–207). NF-kB and Jagged can activate each other (189,208), thereby forming a self-perpetuating loop that maintains both high levels of NF-kB and Notch-Jagged signaling. NF-kB can promote the likelihood for hybrid E/M cells to gain stemness (140), and as discussed above, Notch-Jagged signaling can stabilize cells in a hybrid E/M phenotype; therefore post-therapy inflammatory conditions can promote a drug-resistant subpopulation that can be in hybrid E/M phenotype (Figure 9).

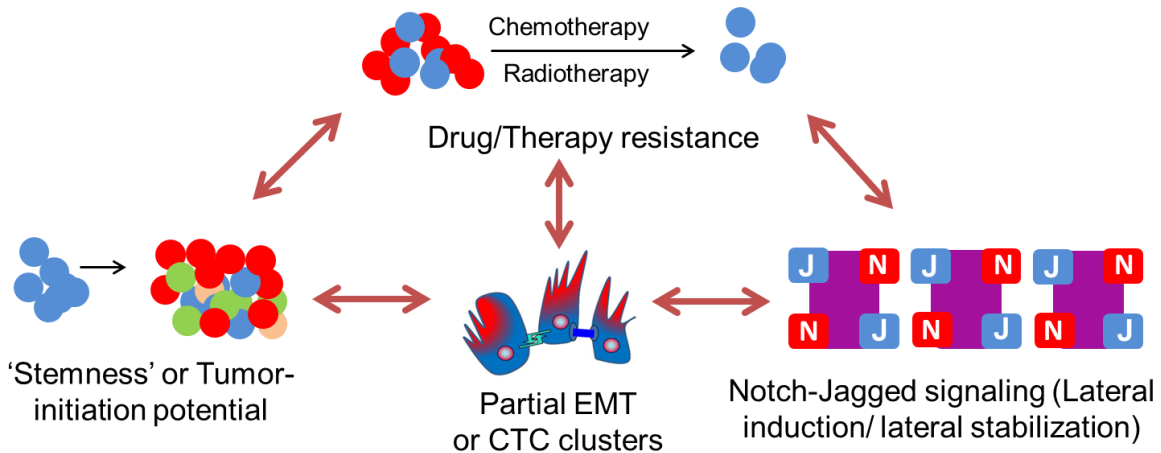

*Figure 9 Interplay between Notch-Jagged signaling, Partial EMT or CTC clusters, stemness, and therapy/drug resistance. Cells in a hybrid E/M or partial EMT phenotype (as present in CTC clusters) can possess a much higher tumor-initiation potential ('stemness') and drug resistance as compared to a completely mesenchymal phenotype. These cells can maintain their 'metastable' hybrid E/M phenotype via Notch-Jagged signaling that promotes lateral stabilization (maintenance of same cell fate in neighboring cells) and/or lateral induction (propagation of the same fate as of its own to the neighbor) among a population of cells. This lateral induction can also be utilized to propagate drug resistance among a small subpopulation of cells referred to as 'Cancer Stem Cells' (CSCs).*

**Conclusion**

*Partial EMT: Primary 'bad actors' of metastases*

EMT is a fundamental process in embryonic development and tissue repair that is aberrantly activated during the progression of cancer and fibrosis. Multiple cycles of EMT and MET are involved in organogenesis but usually not during adult homeostasis (3,4). EMT was first described as "epithelial-mesenchymal transformation" in the pioneering work by Elizabeth Hay on primitive streak formation in the chick (209), however, later the term "transformation" was replaced with "transition" with the evidence accumulating that EMT was different from neoplastic transformation and that it was a reversible process. Recently, the term "transition" is giving way to "plasticity" with an increasing appreciation of the notion that EMT is not an 'all-or-none' response, rather involves intermediate state(s) with important functional consequences in cancer metastasis as well as drug resistance and subsequent tumor relapse (4,210–212).

The appreciation of EMT not being an 'all-or-none' process is relatively recent in EMT associated with cancer and fibrosis, but has been generally accepted in wound healing and collective cell migration during embryonic development, especially gastrulation, neural crest migration and branching morphogenesis (7,8,95,213,214). In cancer-related EMT, the concept of partial or incomplete EMT was initially proposed to reconcile the paradox that despite a presumed role of EMT in cancer progression, most metastatic carcinomas had well-differentiated epithelial characteristics; and it was difficult to identify cells having undergone EMT within the carcinoma tissue *in vivo* (2). Recent experimental evidence about cancer cells in primary tumor, cell lines as well as in circulation (CTCs) (13,63,103,215) have bolstered this concept and has moved it to a focal point in the EMT research.

Cells in a partial or intermediate EMT phenotype are likely to score multiple advantages over cells that have completed EMT or crossed the full mesenchymal 'tipping point'. First, these cells can garner advantages specific to collective sheet or cluster migration – such a migration obviates the need for all cells to respond to external chemotactic signals, allowing for the passive migration of many carcinoma cells, and underlying the unexpected association of E-cadherin with tumor aggression (6,216). Second, these cells display sufficient plasticity to switch to enable a switch back to colonization, yet primed for subsequent metastatic rounds (40,217). Third, these cells are likely to be clustered together in the blood and are therefore anoikis-resistant, an essential trait for efficient metastasis (15). Also, clusters have a greater chance to get trapped in narrow blood vessels, therefore favoring extravasation into distant organs (14). Fourth, these cells can be immune-resistant and chemo-tolerant; and can even be enriched in the population following many therapy-related stresses such as inflammation and radiation (65,218,219). Fifth, these cells can have a much higher (~50-times) tumor-initiating and metastatic potential than cells in complete EMT phenotype (15,16,116,123,133,140). Sixth, due to their residual cell-cell adhesion, these cells might form clusters of CTCs with other cell types such as leukocytes and fibroblasts and/or maintain the clusters via Notch-Jagged signaling among themselves, thereby harnessing their 'ecological engineering' skills during circulation (165). Collectively, the cells in a partial EMT or hybrid E/M phenotype have a much large repertoire of survival strategies in all stress conditions – be it shear stress in circulation, or stress due to therapeutic assaults; and are therefore better armed to seed metastases at distant organs and coordinate tumor relapse. Not surprisingly, these cells are being increasingly observed in many aggressive malignancies (104,199,220–229), strongly suggesting that partial EMT phenomena are more likely to happen *in vivo* than complete EMT (169,230). Isolating CTC clusters (231) and testing them for partial EMT characteristics might be the most promising diagnostic approach in the clinic.

## Acknowledgements

*We have benefited from useful discussions with Mary C. Farach-Carson, Donald S. Coffey, Samir A. Hanash, Kenneth J. Pienta, Ilan Tsarfaty, and Sendurai A. Mani. This work was supported by National Science Foundation (NSF) Center for Theoretical Biological Physics (NSF PHY-1427654) and NSF grant MCB-1214457. HL and JNO are also supported as CPRIT (Cancer Prevention and Research Institute of Texas) Scholar in Cancer Research of the State of Texas at Rice University. EB-J was also supported by a grant from the Tauber Family Funds and the Maguy-Glass Chair in Physics of Complex Systems. MB was also supported by FAPESP (Grant 2013/14438-8). ML*

*was also supported by a training fellowship from Keck Center for Interdisciplinary Bioscience Training of the Gulf Coast Consortia (CPRIT Grant RP140113).*